\documentclass[pre,twocolumn,aps,showpacs,superscriptaddress]{revtex4}
\usepackage{graphics}
\usepackage{epsfig}
\usepackage{amsmath}
\usepackage{amssymb}
\usepackage{bm}
\usepackage{color}

\begin{document}

\title{Critical viscoelastic response in jammed solids}
\author{Brian P. Tighe}
\affiliation{Delft University of Technology, Process \& Energy Laboratory, Leeghwaterstraat 44, 2628 CA Delft, The Netherlands}

\date{\today}

\begin{abstract}
We determine the linear viscoelastic response of jammed packings  of athermal repulsive viscous spheres, a model for emulsions, wet foams, and soft colloidal suspensions. We numerically measure the complex shear modulus, a fundamental characterization of the response, and demonstrate that low frequency response displays dynamic critical scaling near unjamming. Viscoelastic shear response is governed by the relaxational eigenmodes of a packing. We use scaling arguments to explain the distribution of eigenrates, which develops a divergence at unjamming. We then derive the critical exponents characterizing response, including a vanishing shear modulus, diverging viscosity, and critical shear thinning regime. Finally, we demonstrate that macroscopic rheology is sensitive to details of the local viscous force law. By varying the ratio of normal and tangential damping coefficients, we identify and explain a qualitative difference between systems with strong and weak damping of sliding motion. When sliding is weakly damped there is no diverging time scale, no diverging viscosity, and no critical shear thinning regime.
\end{abstract}
\pacs{83.60.Bc,63.50.-x,64.60.Ht}

\maketitle

\begin{figure}[tbp]
\centering
\includegraphics[clip,width=0.95\linewidth]{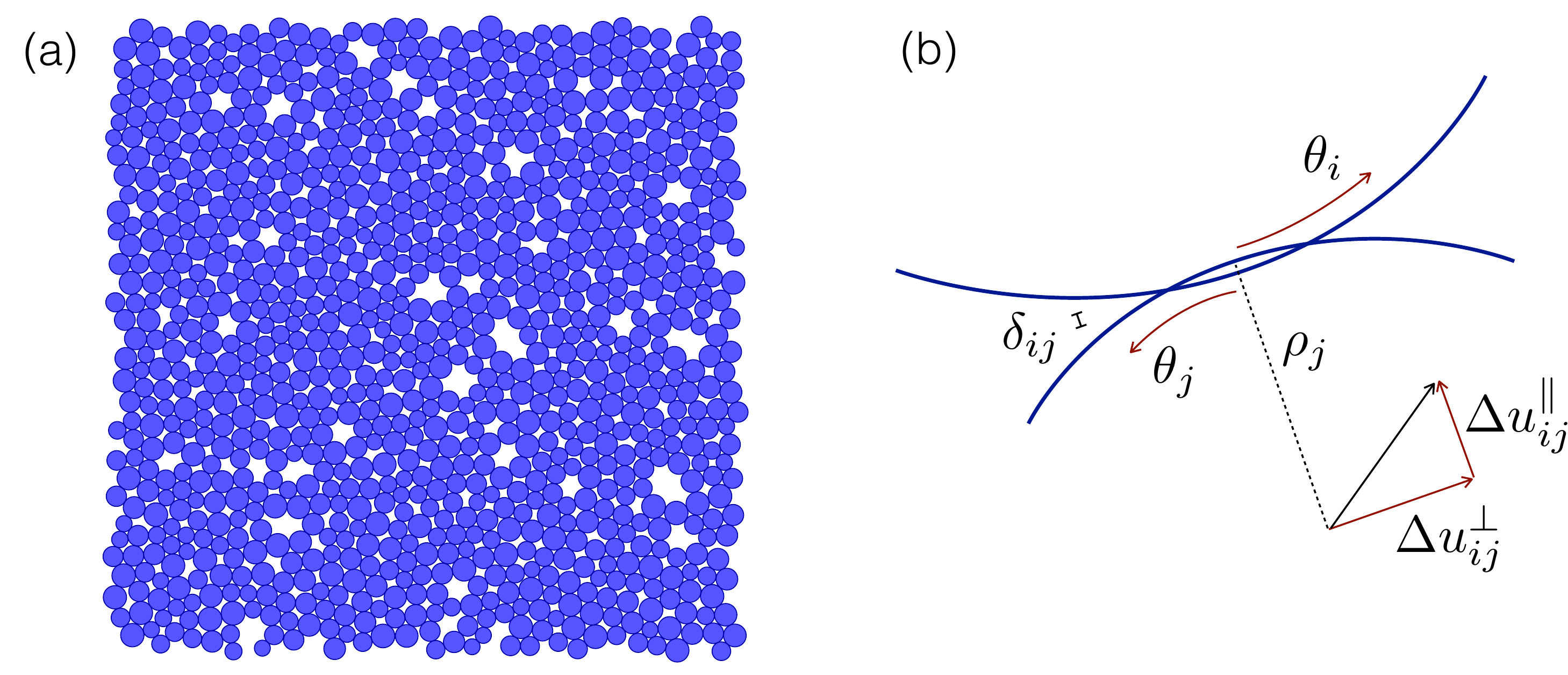}
\caption{(a) Packing of soft repulsive spheres close to unjamming. Particles with no force-bearing contacts (``rattlers'') have been removed.  The mean coordination is $z = 4.03$, close to the isostatic bound $z_c = 4$. (b) Relative motion of two disks in the center of mass frame of the upper disk. Particles exert an elastic force proportional to their overlap $\delta_{ij}$. Viscous forces oppose the relative normal velocity $\Delta \dot u^\parallel_{ij}$ and relative tangential velocity $\Delta \dot u_{ij}^\perp - (\rho_i \dot \theta_i + \rho_j \dot \theta_j)$ at the contact.}
\label{fig:relmotion}
\end{figure}

Sufficiently dense packings of droplets in an emulsion, bubbles in a foam, or soft particles in a colloidal suspension form a jammed, mechanically rigid state
 \cite{princen86,clusel09,katgert10b,nordstrom10b}. Close to the (un)jamming point, which occurs at a critical volume fraction $\phi_c \approx 0.84$ and $0.64$ in $d = 2$ and $3$ dimensions \cite{bolton90,durian95,ohern03,vanhecke10}, contact deformations are small and  particles are nearly spherical. Hence near unjamming these amorphous, viscoelastic solids can be modeled as athermal packings of soft spheres interacting via elastic and viscous contact forces \cite{durian95}, as in Fig.~\ref{fig:relmotion}a.

One noted hallmark of the unjamming transition is the scaling of the quasistatic shear modulus $G_0$, which vanishes on approach to $\phi_c$ as $G_0/k \sim (\phi - \phi_c)^{1/2}$, when measured in units of the microscopic contact stiffness $k$ \cite{ohern03}
 \footnote{The stiffness $k_{ij}$ of the contact between particles $i$ and $j$ is the second derivative of the elastic pair potential $V(r_{ij})$, where $r_{ij}$ is the distance between the particles' centers. }. This anomalous scaling reflects the inherent softness of materials close to unjamming and presages their loss of rigidity.
 Though unquestionably present in numerics \cite{ohern03,ellenbroek06,agnolin07c,zaccone11,tighe11,dagois-bohy12}, the scaling of $G_0$ has yet to be observed experimentally; in fact there have been few experimental results to date that unambiguously confirm scaling relations emerging from the jamming paradigm \cite{katgert10b}. 
 This is at least partly due to the idealized nature of the frictionless soft sphere model; though granular matter is a common testbed for jamming \cite{majmudar07,lechenault10,metayer11},  friction qualitatively alters the way materials jam \cite{somfai07,shundyak07,henkes10}. Foams and emulsions are closer realizations of frictionless soft spheres but present their own experimental challenges; quasistatic measurements, for example, are sensitive to coarsening and other sorts of aging on long time scales \cite{hohler05}.

The best opportunity to confront  theoretical and experimental studies of jammed viscoelastic solids may reside in their {\em finite rate} response \cite{roux05,olsson07,roux08,hatano08,hatano09,otsuki09,xu09,wyart10,tighe10c,tighe11,seth11,sexton11,gomez12}, which is experimentally accessible and physically significant in its own right \cite{liu96,mason97,cohen-addad98,gopal03,katgert08,kropka10,krishan10,nordstrom10,mobius10,lietor-santos11}.
Linear viscoelastic response \cite{hatano09,tighe11} introduces a finite rate while keeping the strain amplitude infinitesimal; it  bridges the gap between linear quasistatic response and response at finite strain and rate, including steady flows \cite{durian95,roux05,olsson07,roux08,hatano08,otsuki09,tighe10c,seth11,sexton11} and shocks \cite{gomez12}. 
Here we measure numerically and calculate analytically the complex shear modulus of jammed viscous sphere packings,
\begin{equation}
G^*(\omega) = G'(\omega) + \imath G''(\omega) \equiv \frac{\sigma(\omega)}{\gamma(\omega)} \,.
\label{eqn:Gstar}
\end{equation}
$G^*(\omega)$ describes the linear response of a material subjected to oscillatory shear at frequency $\omega$, where $\sigma(\omega)$ is the stress amplitude and $\gamma(\omega)$ is the strain amplitude. 
Its real and imaginary parts, $G'$ and $G''$, are respectively known as the storage and loss modulus, while the zero frequency limits of $G'$ and $G''/\omega$ give the quasistatic shear modulus $G_0$ and dynamic viscosity $\eta_0$. 
 
 \begin{figure}[tbp]
\centering
\includegraphics[clip,width=0.99\linewidth]{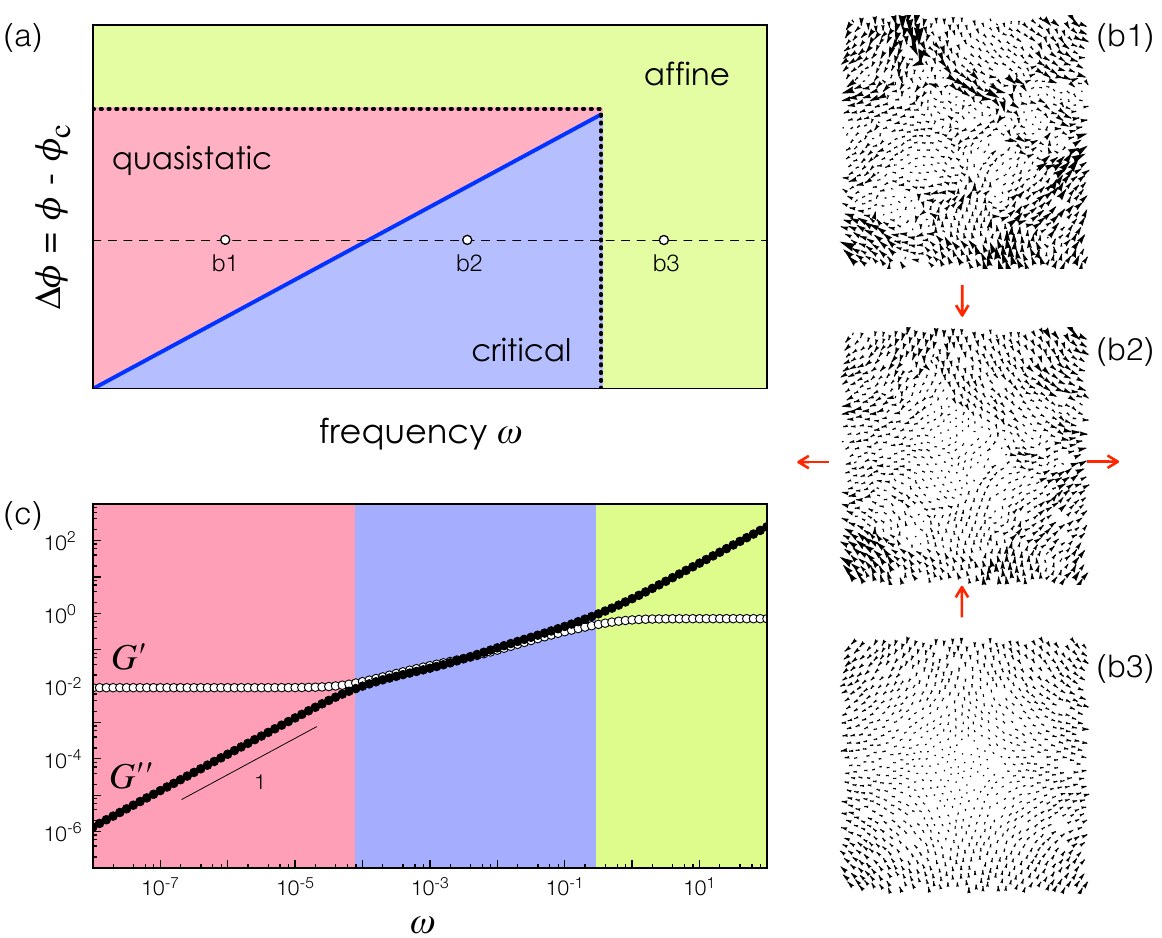}
\caption{(a) Linear viscoelastic response is governed by two control parameters, the driving frequency $\omega$ and the excess packing fraction $\Delta \phi$ or, interchangeably, the excess coordination $\Delta z$  or pressure $p$. There are three qualitatively distinct regimes of response distinguished by the degree of non-affinity of the particles' displacements (b1-b3;  arrows indicate the direction and magnitude of a particle's motion under the pure shear depicted in b2), and the scaling of the complex shear modulus (c), shown here for the packing in Fig.~\ref{fig:relmotion}a. }
\label{fig:regimes}
\end{figure}

This work builds on and substantially broadens results reported in Ref.~\cite{tighe11}. We  demonstrate that viscoelastic response near unjamming is fundamentally related to the relaxational modes of overdamped packings, and that the complex shear modulus is ultimately determined by the density of states $D(s)$ of relaxational eigenrates $s$. We show that displacements at low frequency are consistent with quasistatic response when $\omega \tau^* \ll 1$, with a crossover time scale
\begin{equation}
\tau^* \sim \frac{\tau_0}{\phi - \phi_c} 
\label{eqn:crossover}
\end{equation}
that diverges at unjamming when sliding contacts strongly dissipate energy.
The microscopic time scale $\tau_0$ is determined by coefficients in the viscous and elastic force laws. Because $\tau^*$ diverges at $\phi_c$,  there is no quasistatic response at the unjamming transition. 

For times shorter than $\tau^*$ the dynamics enter a critical rate-dependent regime in which
\begin{equation}
G^* \sim k \,(\imath \omega \tau_0)^\Delta 
\label{eqn:shearthinning}
\end{equation}
with $\Delta  = 1/2$.
Equivalent $1/t^\Delta$ decay of the shear relaxation modulus $G_r(t)$ has previously been observed in simulations of athermal suspensions close to jamming \cite{hatano09}.
We refer to this as the {\em critical} or {\em shear thinning} regime, because doubling the driving rate increases the shear stress by less than a factor of two \footnote{Conventionally the term ``shear thinning'' is reserved for steady flow, but we are not aware of a suitable analog for oscillatory rheology.}.
At high frequencies $\omega \tau_0 \gg 1$, particle displacements smoothly follow the deformation gradient and the storage and loss moduli reflect the microscopic stiffness and damping coefficients. These {\em quasistatic}, {\em critical}, and {\em affine} regimes of response are summarized in Fig.~\ref{fig:regimes}.

The  anomalous quasistatic response near unjamming has its origins in the strongly {\em non-affine} nature of deformations in jammed packings \cite{vanhecke10}. Non-affine motion occurs when individual particles do not smoothly follow the macroscopic deformation gradient; see Fig.~\ref{fig:regimes}b. Non-affinity manifests microscopically as an anomalous abundance of relative tangential motion at contacts, i.e.~sliding, over relative normal motion \cite{ellenbroek06,wyart08}. The preference for sliding is closely linked to the character of {\em floppy modes}, zero energy deformations with no relative normal motion \cite{alexander,tkachenko99,roux00,vanhecke10}. While true floppy modes only exist in the unjammed phase, low energy excitations in the jammed state, which dominate quasistatic deformation, resemble floppy modes insofar as they have small normal motions and large sliding motions \cite{wyart05}.

We will show that in viscoelastic response, non-affinity develops in time; the slower the deformation, the stronger the dominance of sliding motion. The three regimes of response are characterized by different degrees of non-affinity, as illustrated in Fig.~\ref{fig:regimes}a,b. For a given packing fraction, non-affinity is maximal in the quasistatic limit, while in the affine regime sliding and normal motions are comparable. These two limits are connected by the critical regime, where non-affine motion is present but not fully developed.

Because sliding plays a dominant role in low frequency response, the loss modulus is sensitive to the microscopic coupling between dissipation and sliding. We will show that a qualitatively different rheology results when sliding is weakly damped. Hence details of the microscopic interactions, in this case the local viscous force law, can dramatically alter the global rheology -- in sharp contrast to the usual scenario for critical scaling in equilibrium systems \cite{hohenberghalperin}.

\section{Linear viscoelastic response of soft spheres}
\label{sec:formalism}

We begin by developing the equations of motion describing linear viscoelastic response. Microscopic elastic and viscous interactions are linearized about a numerically-generated reference configuration, in this case a static soft sphere packing such as the one in Fig.~\ref{fig:relmotion}a. To make numerical measurements, we average over ensembles of packings prepared at different confining pressures $p$. 
Average properties such as the complex shear modulus $G^*(\omega)$ and the relaxational density of states $D(s)$ have characteristic features that depend only on the distance to jamming, e.g.~the pressure $p$ or the excess volume fraction $\Delta \phi := \phi - \phi_c$. We will determine these features using scaling arguments. While our numerical methods require computer-generated packings as input, the predicted scaling relations depend only on global parameters such as the volume fraction or pressure.

\subsection{Equations of motion}
When a packing in $d = 2$ dimensions is deformed, the particles undergo displacements $\lbrace \vec u_i \rbrace_{i = 1 \ldots N}$ from their equilibrium positions. We consider viscous forces capable of applying torques, which couple displacements and rotations  $\lbrace \theta_i \rbrace_{i = 1 \ldots N}$. For macroscopic deformations, the packing also undergoes some pure shear strain $\gamma$ (Fig.~\ref{fig:regimes}b).  The deformation is thus characterized by $3N + 1$ degrees of freedom: the vector components of the displacement, the particle rotations, and the strain $\gamma$. Although we present numerical results for two dimensions, all known critical exponents characterizing jamming are independent of dimension \cite{vanhecke10}. We will show in Sec.~\ref{sec:response} that this is true for viscoelastic response, as well.

We study soft particles interacting via repulsive elastic  contact forces; this is the canonical model system for the study of (un)jamming \cite{ohern03,vanhecke10}. The elastic force between disks of radii $\rho_i$ and $\rho_j$ is linearly proportional to their overlap $\delta_{ij} = \rho_{i} + \rho_j -  \Delta r_{ij}$, where $\Delta r_{ij}$ is the center-to-center distance. Overlapping disks also experience viscous forces that oppose relative normal and tangential motions at the contact; see Fig.~\ref{fig:relmotion}b. We are interested in slowly driven systems, for which inertial effects can be neglected. Setting the particle masses to zero yields overdamped equations of motion. Overdamped soft spheres are often referred to as the ``bubble model'' after Durian, who introduced them as a model for wet foams \cite{durian95}. 

The Lagrangian equations of motion for the bubble model are
\begin{equation}
\sum_i \left( \frac{\partial U}{\partial q_i} + \frac{\partial R}{\partial \dot Q_i} \right) = F_i \,.
\label{eqn:lagrangeeom}
\end{equation}
Here the $\lbrace Q_i \rbrace_{i = 1 \ldots 3N+1}$ label all the degrees of freedom of the system: the vector components of the particle displacements, the particle rotations, and the shear strain.  $U$ is the elastic potential energy and $R$ is the Rayleigh dissipation function, both defined below. $F_i$ is the generalized force conjugate to the degree of freedom $Q_i$. The generalized forces conjugate to the the displacements and rotations represent body forces and couples, respectively; they are all zero. The generalized force conjugate to the shear strain is the force moment $\sigma \Omega$, where $\sigma$ is the shear stress and $\Omega$ is the volume of the unit cell.

We linearize the equations of motion (\ref{eqn:lagrangeeom}) about a static packing, while distinguishing the contributions of relative normal and tangential motions \cite{alexander}. To quadratic order in $\gamma$ the elastic potential energy $U$ is
\footnote{
In general, Eq.~(\ref{eqn:V}) also has a term $\sigma^0 \Omega \gamma$ that is linear in the strain $\gamma$. However, we consider only isotropic reference states, for which the shear stress $\sigma^0 = 0$.}
\begin{equation}
U - U^0 =
 \frac{1}{2} \sum_{\langle ij \rangle}\left[
k_{ij} (\Delta u_{ij}^\parallel)^2 - \frac{f_{ij}^0}{\Delta r_{ij}^0}(\Delta u_{ij}^\perp)^2
\right]
 \,.
\label{eqn:V}
\end{equation}
Here $\Delta {\vec u}^\parallel_{ij} = (\vec u_j - \vec u_i) \cdot \hat n_{ij}$ is the relative motion of two  disks along their contact normal $\hat n_{ij}$, and $\Delta {\vec u}_{ij} = \Delta {\vec u}_{ij} - \Delta {\vec u}^\parallel_{ij}$ is the relative tangential motion; see Fig.~\ref{fig:relmotion}b. We take the contact stiffness $k_{ij} \equiv k$ to be the same for every contact. 
Quantities evaluated in the reference state are labeled with a superscript $0$, and $f^0_{ij}$ is the force carried by contact $(ij)$ in the reference state. The contribution proportional to these forces is known as the pre-stress term. 

The dissipation function is  \cite{tighe11}
\begin{equation}
R = \frac{1}{2} \sum_{\langle ij \rangle} \left[
b^\parallel (\Delta \dot u^\parallel_{ij})^2
+ b^\perp 
(\Delta \dot u^\perp_{ij} - \rho_i {\dot \theta_i} - \rho_j {\dot \theta_j})^2 
\right]\,.
\label{eqn:R}
\end{equation}
One can verify that derivatives of $R$ with respect to the particle velocities yield viscous forces damping normal and sliding motion at the contact with damping coefficients $b^\parallel$ and $b^\perp$, respectively. 
Note that two overlapping but otherwise isolated particles separate with a relaxation rate $k/b^\parallel \equiv 1/\tau_0$. In the following we shall express the damping coefficients in terms of the microscopic time scale $\tau_0$:  $b^\parallel \equiv k \tau_0$ and $b^\perp \equiv \beta k \tau_0$. The dimensionless quantity $\beta$ is  the ratio of tangential to normal damping. 

Collecting all degrees of freedom in a vector $|Q(t)\rangle$, the linearized equations of motion can be expressed as a matrix equation,
\begin{equation}
\hat {\cal K} |Q(t) \rangle + \hat {\cal B} |\dot Q(t) \rangle  = \sigma(t) \, |\hat \sigma\rangle \,,
\label{eqn:eom}
\end{equation}
where  $|\hat \sigma \rangle \equiv |\lbrace 0 \rbrace,\Omega \rangle$.
The stiffness matrix $\hat {\cal K}$ and damping matrix $\hat {\cal B}$ capture the elastic and viscous interactions of the particles,
\begin{equation}
{\cal K}_{mn} = \frac{\partial^2 U}{\partial Q_m \, \partial Q_n} \,\,\,
{\rm and} \,\,\,
{\cal B}_{mn} = \frac{\partial^2 R}{\partial \dot Q_m  \, \partial \dot Q_n} \,.
\label{eqn:KandB}
\end{equation}

\subsection{A note on units and scaling}
The stiffness matrix  $\hat {\cal K}$ has units of the microscopic stiffness coefficient $k$, and the damping matrix $\hat {\cal B}$ has units of the damping coefficient $k \tau_0$. In many physical systems, $k$ and/or $k \tau_0$ depend on the amount of overlap $\delta$ between contacting particles \cite{ramirez99,seth11}. 
The typical overlap itself vanishes at unjamming: it scales as $\delta \sim \Delta \phi$ \cite{vanhecke10}. Thus macroscopic observables can inherit a ``trivial'' dependence on the distance to unjamming directly from the microscopic coefficients. In the Hertz contact problem, for example, the pair potential $V \propto \delta^{\alpha}$ with $\alpha = 5/2$, so the  the contact stiffness $k = V'' \propto \delta^{1/2}$ \cite{ramirez99}. As noted above, $G_0/k \sim \Delta \phi^{1/2}$ near unjamming. Hence $G_0 \sim \Delta \phi$ in a system of particles interacting via Hertzian potentials, which is different from the square root scaling observed for harmonic interactions ($\alpha = 2$).
 In our numerics, $k$ and $\tau_0$ are always set to unity; nevertheless, for clarity and generality, we explicitly indicate the $k$ and $\tau_0$ dependence of scaling relations. We omit dependence on the microscopic length scale $\rho$, the typical particle size, because it does not scale.

As we are interested in scaling near unjamming, we must choose an appropriate measure of distance to the transition. We have already introduced $\Delta \phi$ and $p$, which both vanish at unjamming; below we will also encounter the contact number $z = z_c+ \Delta z$, which characterizes the geometry of a packing and approaches the isostatic value $z_c = 2d$ at unjamming \cite{alexander,vanhecke10,dagois-bohy12}. Each measure has its own context-dependent advantages. We  rescale our numerical measurements with $\Delta z$ because, as shall become apparent below, it is the most fundamental of the three. However, as contact numbers are difficult to determine experimentally, it can also be convenient to couch scaling relations in terms of the dimensionless quantities $\Delta \phi$ or $p/k$. All three quantities are related via well-known scaling relations, $ \Delta z^2 \sim \Delta \phi  \sim p/k$ \cite{vanhecke10}, so scaling in terms of $\Delta z$ can always be rewritten as scaling in $\Delta \phi$ or $p/k$.

\section{Dynamic critical scaling}
\label{sec:scaling}

We have shown that the viscoelastic shear response of a soft sphere packing obeys Eq.~(\ref{eqn:eom}). We will now use this equation of motion to determine the response to oscillatory driving at finite frequency. 
For a small oscillatory shear stress
$\sigma = \sigma(\omega) e^{\imath \omega t}$ with frequency $\omega$,
the particles oscillate with the same frequency,
$|Q(t)\rangle = e^{\imath \omega t} \, |Q(\omega) \rangle$. In general the response is out of phase with the stress.
The equation of motion for oscillatory rheology in overdamped soft spheres is then
\begin{equation}
\hat {\cal K} |Q(\omega) \rangle + \imath \omega \hat {\cal B} |Q(\omega) \rangle  = \sigma(\omega) \, |\hat \sigma \rangle \,.
\label{eqn:osc_eom}
\end{equation}

Eq.~(\ref{eqn:osc_eom})  can be solved numerically for $|Q(\omega)\rangle$. Of particular interest is the complex shear modulus, $G^*(\omega) = \sigma(\omega)/\gamma(\omega)$, which provides a fundamental characterization of the shear response of the packing. The frequency dependent shear strain $\gamma(\omega)$ can simply be read off from $|Q(\omega)\rangle$, namely $\gamma(\omega) = \Omega^{-1}\langle \hat \sigma |Q(\omega) \rangle$.

\subsection{Numerical results}

\begin{figure*}[tbp]
\centering
\includegraphics[clip,width=0.45\linewidth]{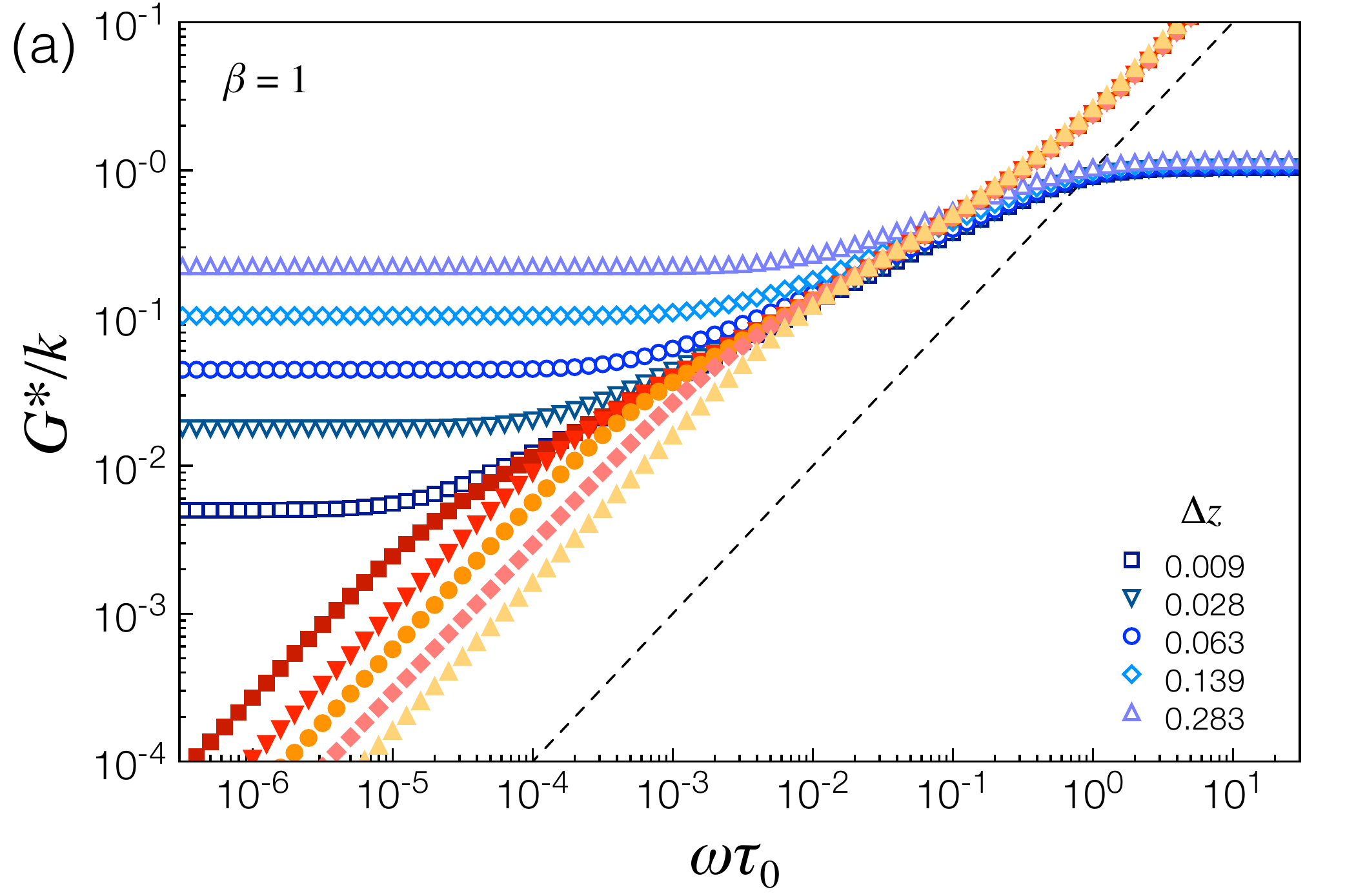}
\hspace{0.05\linewidth}
\includegraphics[clip,width=0.45\linewidth]{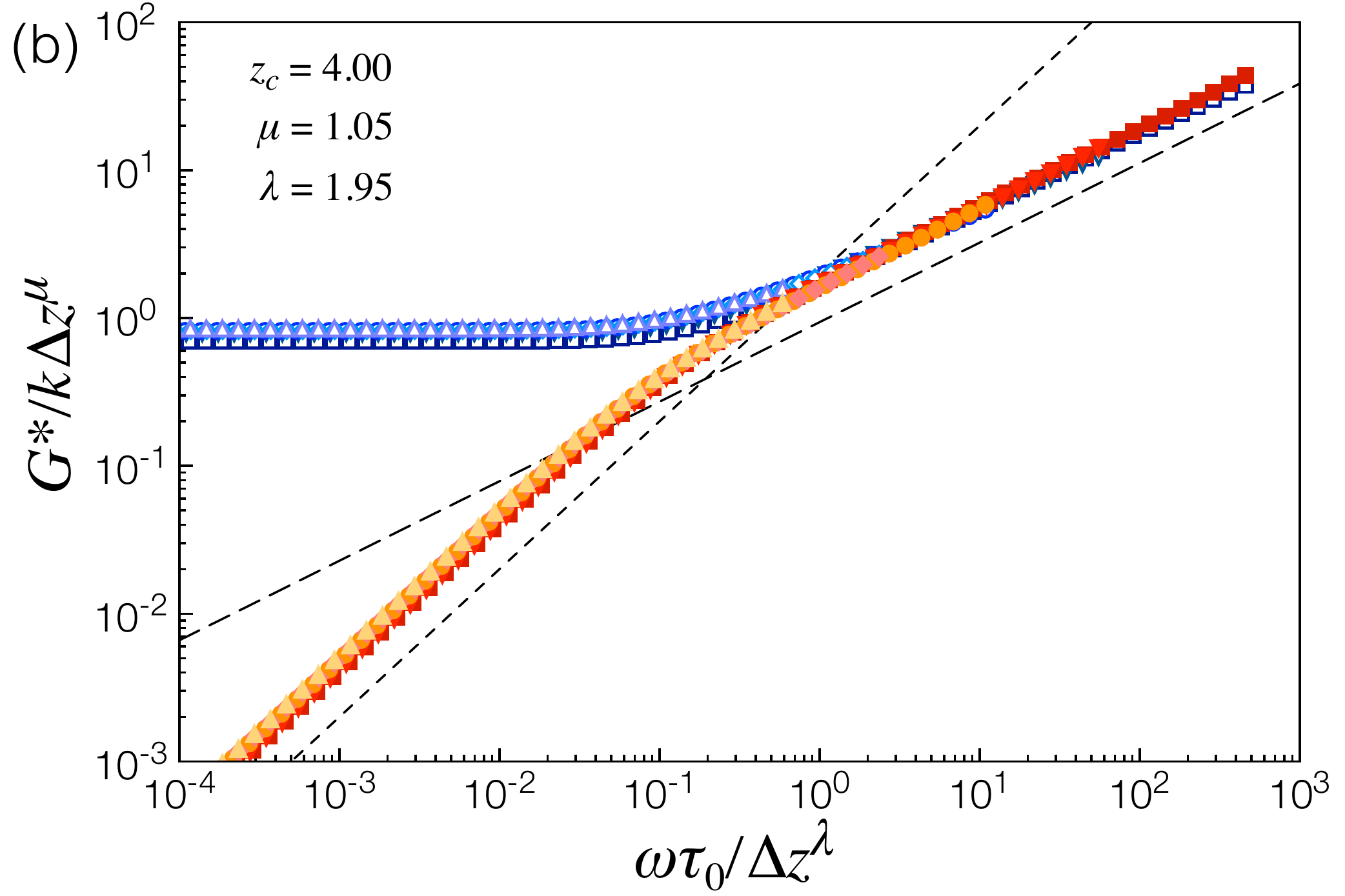}
\caption{
(a) Storage modulus $G'$ (open symbols) and loss modulus $G''$ (filled symbols) for a range of excess coordination $\Delta z$ (legend) close to unjamming. Normal and tangential damping coefficients are equal ($\beta = 1$). The dashed line has slope 1.
(b) Critical collapse of the low frequency data in (a). The short dashed line has slope $1$; the long dashed line has slope $\Delta = \mu/\lambda = 0.54$.
} 
\label{fig:Gstar_rel}
\end{figure*}

We evaluate the complex shear modulus numerically in an ensemble of packings generated with the molecular dynamics protocol of Somfai et al.~\cite{somfai05}. Each packing contains $N = 1024$ weakly polydisperse particles prior to the removal of ``rattlers,'' i.e.~non-force bearing particles. For each excess contact number $\Delta z$ the response is averaged over $\approx 50$ packings. Here and below, the pre-stress term of the elastic energy $U$ has been neglected when evaluating the stiffness matrix $\hat {\cal K}$; this approximation is justfied in Section \ref{sec:response}. We find similar results when the pre-stress term is included, albeit with greater scatter.

The complex shear modulus for varying excess coordination $\Delta z$ is plotted in Fig.~\ref{fig:Gstar_rel}a. To interpret the figure, it is helpful to recall that the simplest possible viscoelastic solid -- the Kelvin-Voigt solid -- has complex shear modulus $G^*_{\rm KV} = G_{\rm KV} + \imath \eta_{\rm KV} \omega$, i.e.~its storage modulus is flat and its loss modulus is linear in frequency.
For both low and high frequencies, the packings of Fig.~\ref{fig:Gstar_rel}a indeed have flat storage moduli. At high frequencies the storage modulus is consistent with the microscopic spring constant, $G'/k \sim {O}(1)$. The low frequency plateau in $G'$, however, does not match the high frequency plateau; the  storage modulus is {\em softer} at low frequencies than at high, and therefore softer than one would na\"ively anticipate based on the contact stiffness. This mismatch grows as the unjamming transition is approached: the height of the low frequency plateau diminishes. 

The low and high frequency plateaus in the storage modulus are connected by a range of apparent power law scaling over a frequency interval of finite width. The power law is sublinear, and therefore the material is {\em shear thinning}. The upper bound of the shear thinning regime occurs for $\omega\tau_0  \sim {O}(1)$, independent of the distance to jamming. In contrast, the lower bound decreases -- and the shear thinning regime grows wider -- as unjamming  is approached.

Similar trends are observed in the loss modulus. Like the Kelvin-Voigt material, $G''$ is linear at low and high frequencies. However, it also displays an anomalous power law regime for intermediate frequencies. The apparent exponent is less than one and similar to that observed in the storage modulus. The upper and lower bounds of the shear thinning regime coincide  with those observed in the storage modulus.

\subsection{Scaling}
We noted above that the quasistatic ($\omega \rightarrow 0$) shear modulus $G_0$ of soft sphere packings scales as
\begin{equation}
\frac{G_0}{k} \sim 
\Delta \phi^{\mu/2} \sim \Delta z^{\mu} \,,
\label{eqn:G0}
\end{equation}
with $\mu \approx 1.0$ \cite{ohern03,dagois-bohy12}. Thus $G_0$ vanishes continuously at the unjamming transition. In many soft matter systems with a rigidity transition, finite frequency response near the transition displays dynamic critical scaling \cite{winter86,hohenberghalperin}. We now show that this is also true of viscoelastic solids near unjamming. First we demonstrate scaling numerically; then, in Section \ref{sec:response}, we explain its origins.

We begin by making the scaling ansatz 
\begin{equation}
\frac{1}{k}G^*(\omega,\Delta z) = \Delta z^{\mu} \, {\cal G}^*\left(\frac{\omega \tau_0}{\Delta z^{\lambda}}  \right) \,,
\end{equation}
where the scaling function ${\cal G}^*(x) \equiv {\cal G}'(x) + \imath {\cal G}''(x)$ obeys
\begin{equation}
{\cal G}'(x) \sim \left \lbrace
  \begin{array}{cl}
  {\rm const} & x \ll 1 \\
  x^\Delta & x \gg 1 \,,
  \end{array} \right.
\end{equation}
and
\begin{equation}
{\cal G}''(x) \sim \left \lbrace
  \begin{array}{cl}
  x & x \ll 1 \\
  x^\Delta & x \gg 1 \,. \end{array} \right.
\end{equation}
The $x \ll 1$ form of the dimensionless scaling function ${\cal G}^*$ is chosen to reflect the frequency dependence of the Kelvin-Voigt material. The same exponent $\Delta$ appears in both ${\cal G}'$ and ${\cal G}''$ for $x \gg 1$; this follows from the Kramers-Kronig relations \footnote{To see this quickly, consider the local Kramers-Kronig relation $G'' \approx (\pi/2)\omega \, \partial_\omega G'$, which is appropriate when the storage modulus is weakly dependent on frequency. More careful consideration of the full (integral, hence non-local) Kramers-Kronig relations leads to the same conclusion.}.
To guarantee that $G^*(\omega > 0)$ remains finite when $\Delta z = 0$, the exponents $\mu$, $\lambda$, and $\Delta$ must obey 
\begin{equation}
\mu = \lambda \Delta \,.
\end{equation}

Dynamical critical scaling relates two qualitatively different regimes of response. The rheology of Fig.~\ref{fig:Gstar_rel}a, however, clearly displays three regimes -- these are the quasistatic, critical, and affine regimes of Fig.~\ref{fig:regimes}. In the following we seek to relate the quasistatic and critical regimes; we therefore exclude frequencies $\omega\tau_0 > 0.3$. To test the critical scaling ansatz, we plot $(G^*/k)/\Delta z^{\mu}$ versus $(\omega \tau_0) /\Delta z^{\lambda}$ and vary $\mu$ and $\lambda$ to find the best data collapse. We find $\mu = 1.05(5)$ and a dynamical exponent  $\lambda = 1.95(5)$. The resulting collapse is excellent, as shown in Fig.~\ref{fig:Gstar_rel}b. The shear thinning exponent $\Delta = \mu/\lambda = 0.54$ is also in good agreement with the data. In Section \ref{sec:scaling} we will predict $\mu = 1$, $\lambda = 2$, and $\Delta = 1/2$, but for now they are phenomenological.

The data collapse in Fig.~\ref{fig:Gstar_rel}b indicates that low frequency dynamics near jamming display dynamic critical scaling. This has several immediate implications. First, the scaling function has a characteristic time scale
\begin{equation}
\tau^* = \frac{\tau_0}{\Delta z^{\lambda}} \,.
\end{equation}
The quasistatic limit corresponds to frequencies $\omega \ll 1/\tau^*$.
In this limit we can read off the quasistatic shear modulus $G_0$ and dynamic viscosity $\eta_0$. The shear modulus is $G_0 \equiv \lim_{\omega \rightarrow 0} G'(\omega) \sim k \, \Delta z^\mu$, which recovers the quasistatic scaling of Eq.~(\ref{eqn:G0}) as expected. The dynamic viscosity is determined by the imaginary part of the scaling function, $\eta_0 \equiv \lim_{\omega \rightarrow 0} G''(\omega)/\omega$. It scales as 
\begin{equation}
\eta_0  \sim  G_0 \tau^* \sim \frac{k \tau_0}{\Delta z^{\lambda - \mu}}\,,
\label{eqn:eta0}
\end{equation}
which is diverging. Note that the viscosity does not diverge with the same exponent as the time scale $\tau^*$ due to the vanishing shear modulus.

The storage and loss moduli are dominated by the quasistatic moduli for frequencies up to $1/\tau^*$. 
 On time scales shorter than $\tau^*$, however, the dynamics enter a qualitatively different, critical regime.
 Instead, both the storage and loss moduli display an anomalous power law scaling with frequency,  expressed succinctly as
\begin{equation}
G^* \sim k \, (\imath  \omega \tau_0)^\Delta \,.
\end{equation}
Because $0 < \Delta < 1$, this response is a form of shear thinning. $\Delta$ is a new critical exponent near unjamming.

Our low frequency scaling relations can be compared to known results calculated in the strictly quasistatc limit, which corresponds to setting $\omega$ to zero in Eq.~(\ref{eqn:osc_eom}). If the limit is continuous, then quasistatic particle trajectories should be a good approximation to response at sufficiently low frequencies, where damping is weak. We noted above that the exponent $\mu$ characterizing the low frequency shear modulus is in good agreement with the established quasistatic value of $1.0$, which suggests the quasistatic limit is indeed continuous. 
Under quasistatic deformation the typical magnitude of sliding motion is known to scale as $\Delta u^\perp \sim \gamma/\Delta z^{1/2}$ \cite{ellenbroek06,wyart08}.  This can be related to the viscosity by balancing the dissipated power $(1/2)\eta_0 (\omega \gamma)^2$ with the microscopic dissipation rate ${\vec f}^{\rm visc} \cdot \Delta \dot {\vec u} \sim k \tau_0 (\omega \, \Delta u^\perp)^2$, giving $\eta_0 \sim  k \tau_0/\Delta z$. 
Note that the quasistatic regime breaks down when $\eta_0 \omega / G_0 \ll 1$ is violated, i.e.~when the quasistatic scaling relations no longer permit damping to be weak compared to storage.
\section{Relaxations and response}
\label{sec:response}

In Section \ref{sec:scaling} we demonstrated numerically that shear response near jamming displays dynamic critical scaling. Simply demonstrating scaling does not explain its origin, so we now seek to identify the physical mechanism underlying packings' critical response.
We begin by showing that viscoelastic response is governed by the relaxational eigenmodes of an overdamped packing. We then numerically determine the relaxational density of states $D(s)$ and introduce a scaling argument to explain the characteristic features of $D(s)$. Finally, we demonstrate that the exponents $\mu$, $\lambda$, and $\Delta$ characterizing oscillatory rheology can be derived from the density of states.

\subsection{Complex compliance and shear modulus}
Anticipating that the eigenmodes of an overdamped packing are evanescent, it is natural to consider the Laplace transform of the equation of motion. In the absence of driving, $\sigma = 0$, this is
\begin{equation}
(\hat {\cal K}+ s_n \, \hat {\cal B}) |s_n \rangle = 0 \,.
\label{eqn:geneigprob}
\end{equation}
Eq.~(\ref{eqn:geneigprob}) is a generalized eigenvalue equation -- ``generalized'' because $\hat{\cal B}$ is not the identity matrix. The equation has eigenvalues $\lbrace s_n \rbrace$ and  orthonormal eigenvectors $\lbrace | s_n \rangle \rbrace$. 

Eq.~(\ref{eqn:geneigprob}) is invariant under rigid body translations and individual particle rotations; thus there are $N+2$ trivial zero modes of the system in two dimensions. A jammed system does not possess non-trivial or ``floppy'' zero modes. The remaining eigenvectors are $\hat {\cal B}$-orthogonal, meaning $\langle s_m | \hat {\cal B}| s_n \rangle = 0$ for $m \neq n$, and  the remaining eigenvalues are negative definite. Physically, the latter property means that the modes are evanescent: a packing deformed along a mode $|s_n\rangle$ relaxes exponentially to the reference configuration with relaxation rate $|s_n|$. For convenience, from this point forward we adopt a sign convention in which the symbol $s_n$ refers to the absolute value of the eigenrate of the $n^{\rm th}$ eigenmode.

We now return to Eq.~(\ref{eqn:osc_eom}) with driving. 
Expressing the response as a superposition of relaxational eigenmodes, one may invert the equation of motion to solve for the complex compliance $J^*(\omega) \equiv \gamma(\omega)/\sigma(\omega)$,
\begin{eqnarray}
{J'(\omega)} &=& \frac{1}{ \Omega} \sum_n 
\frac{1}{\eta_n} \frac{s_n }{\omega^2 +  s_n^2}   \nonumber \\
{J''(\omega)} &=& - \frac{1}{ \Omega} \sum_n 
\frac{1}{\eta_n} \frac{ \omega}{\omega^2 +  s_n^2} \,.
\label{eqn:J}
\end{eqnarray}
We have introduced $\eta_n \equiv {k \langle s_n | \hat {\cal B} | s_n \rangle}/{|\langle s_n | \hat \sigma \rangle |^2}$; the motivation for this nomenclature will become apparent below. Note that the projection  $\langle s_n |\hat \sigma \rangle$ is a measure of how strongly mode $n$ couples to shear. The trivial translational and rotational modes do not couple to shear.

\begin{figure}[tbp]
\centering
\includegraphics[clip,width=0.9\linewidth]{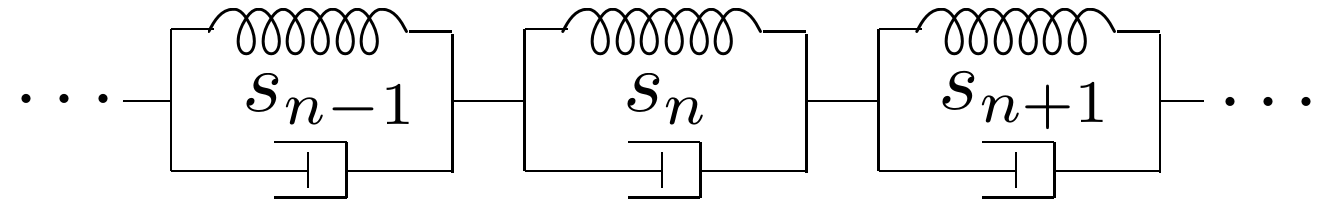}
\caption{Series circuit of Kelvin-Voigt elements. Each element has a characteristic relaxation rate associated with one of a packing's relaxational modes. }
\label{fig:circuit}
\end{figure}

Eq.~(\ref{eqn:J}) completely specifies the microscopic response of the system in terms of properties of the individual modes, including the complex shear modulus  $G^*(\omega) = 1/J^*(\omega)$. It is straightforward to show that response of this form is identical to that of a series circuit of viscoelastic elements, one for each mode, as depicted in Fig.~\ref{fig:circuit}. Such a circuit has a complex shear modulus
\begin{equation}
{G^*} = \left( \sum_n \frac{1}{G^*_n} \right)^{-1} \,,
\end{equation}
Each element has a modulus $G^*_n = G_n + \imath \eta_n \omega$, where $\eta_n$ is defined above and $G_n = \eta_n s_n$.
Hence each Kelvin-Voigt element has a characteristic relaxation rate $G_n / \eta_n = s_n$ equal to the eigenrate of the corresponding mode.
We stress that the series circuit of Fig.~\ref{fig:circuit} is not an {\em ad hoc} model, but a mapping from an exact solution to the equations of motion.

We have now shown that a packing's complex shear modulus is determined by its relaxational modes, and that two quantities are required to quantify each mode: its relaxation rate $s_n$ and the coupling of the $n^{\rm th}$ mode to shear via the mode's ``viscosity'' $\eta_n$. To make progress, we now introduce an approximation: we assume that $\eta_n$ is random, i.e.~independent of $n$ and implicitly of $s_n$ as well, and that it has a typical value $\eta_n \sim k \tau_0$.
We have verified numerically that this is reasonable. 

Under the above assumption, the fact that response can be mapped to the series circuit of Fig.~\ref{fig:circuit} establishes an important intuition. The modes with the lowest eigenrates, i.e.~those that relax slowly, have the smallest effective storage modulus $G_n \sim k \tau_0 s_n$. As a series circuit's low frequency deformation is carried by its softest elements, one infers that low frequency shear will be carried predominantly by the slow modes.

Taking the thermodynamic limit of Eq.~(\ref{eqn:J}) gives
\begin{eqnarray}
kJ'(\omega)  &\sim&  \frac{1}{\tau_0} \int_0^\infty  \frac{s}{s^2+ \omega^2}  D(s) \, {\rm d}s\nonumber \\
kJ''(\omega) &\sim&  - \frac{1}{\tau_0}  \int_0^\infty  \frac{\omega }{s^2+ \omega^2} D(s) \, {\rm d}s  \,.
\label{eqn:compliance}
\end{eqnarray}
The scaling of $J^*$, and hence $G^*$ as well, is now completely governed by the relaxational density of states $D(s)$, which we proceed to determine.

\subsection{Relaxational modes}
\label{sec:DOS}
What are the characteristic features of $D(s)$? And how are they reflected in the response? To answer these questions, we investigate $D(s)$ both numerically and theoretically. We will show that slow modes are both abundant and strongly non-affine.

\begin{figure*}[tbp]
\centering
\includegraphics[clip,width=0.45\linewidth]{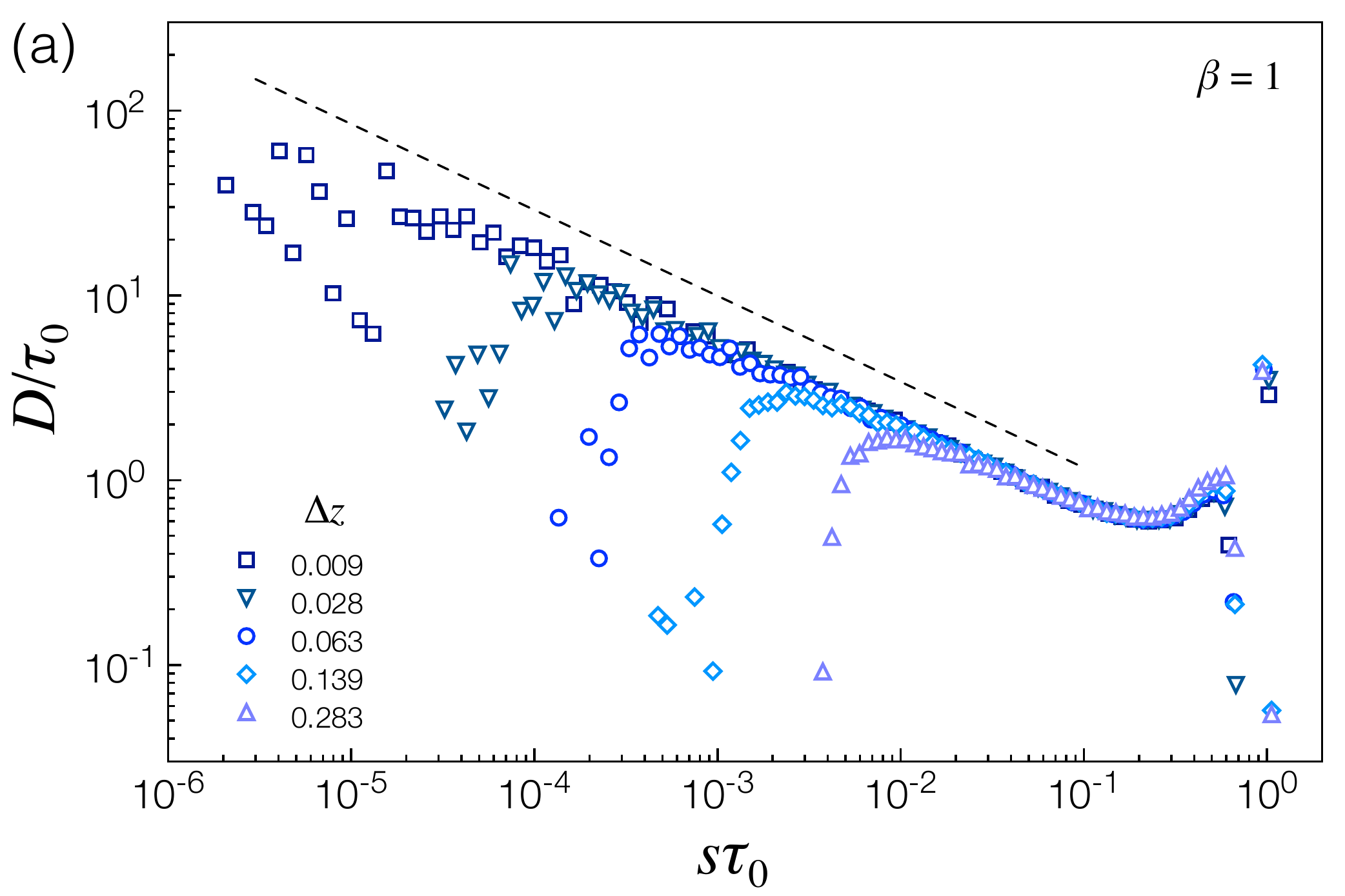}
\hspace{0.05\linewidth}
\includegraphics[clip,width=0.45\linewidth]{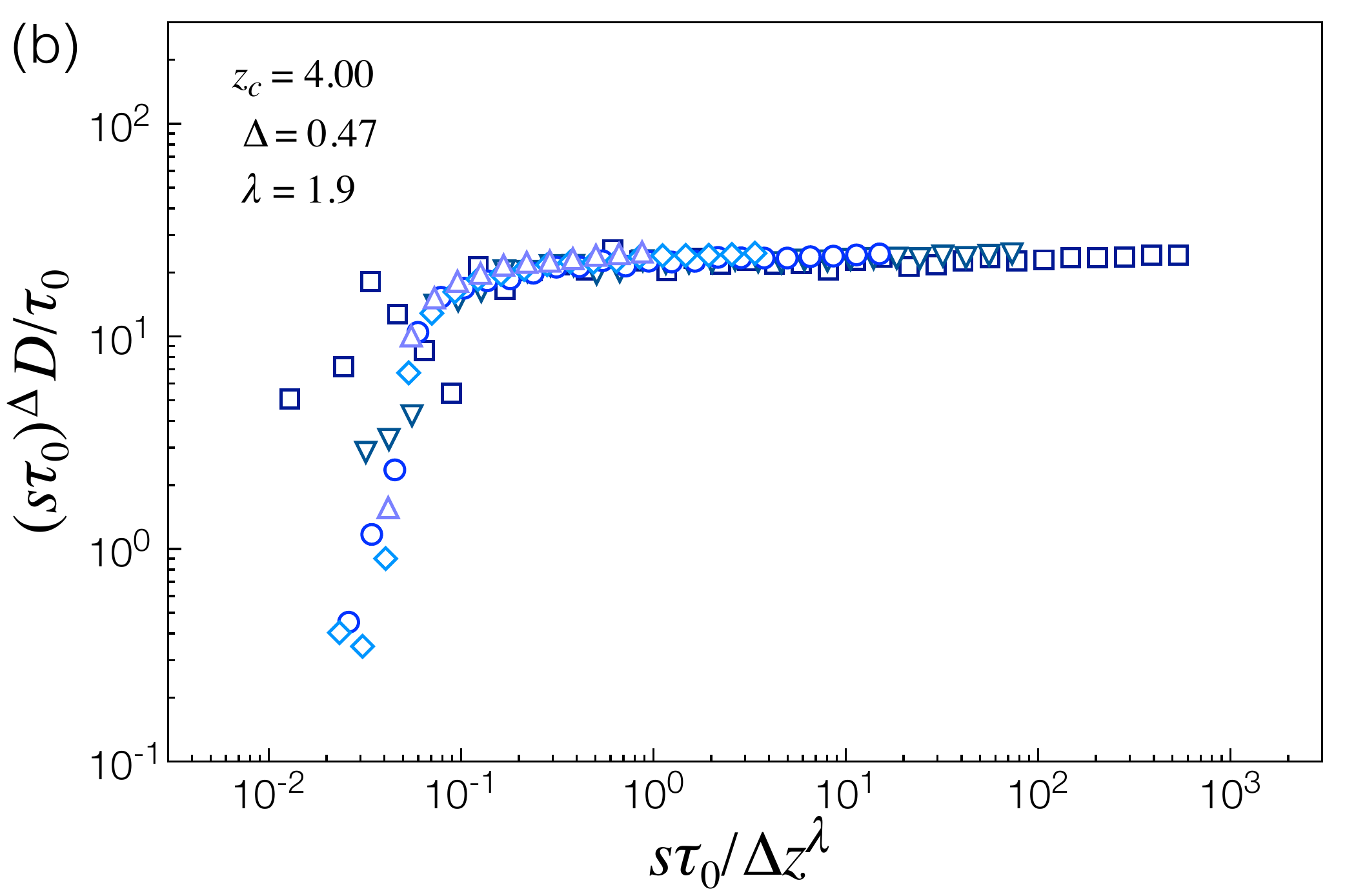}
\caption{
(a) Ensemble averaged relaxational density of states $D(s)$ near the unjamming transition, where the excess coordination $\Delta z$ (legend) vanishes. Normal and tangential damping coefficients are equal ($\beta = 1$). The dashed line has slope $1/2$.
(b) Rescaled data from (a). The exponents $\lambda$ and $\Delta$ have been chosen so as to collapse the crossover at low $s$ and generate a plateau above it.
}
\label{fig:DOS_rel}
\end{figure*}

Fig. ~\ref{fig:DOS_rel}a depicts the relaxational density of states for several different distances to jamming, parameterized by $\Delta z$. Several features stand out. First, the fastest relaxations are on the order of the bare rate $1/\tau_0$ at which two overlapping particles separate. Second, low-$s$ relaxations carry a large statistical weight, and that weight increases as jamming is approached: the closer it is to jamming, the more ways a packing has to relax { slowly}. Third, and more quantitatively, $D(s)$ is approaching a power law divergence  in the limit $\Delta z \rightarrow 0$,
\begin{equation}
\frac{D(s)}{\tau_0} \sim \left(\frac{1}{s \tau_0 }\right)^{\Delta'} \,,
\end{equation} 
for some exponent $\Delta' > 0$.  At finite $\Delta z$ the divergence is cut off at a characteristic rate $s^* \sim \Delta z^{\lambda'}$, for some $\lambda' > 0$. 

To characterize the relaxational density of states, we plot $s^{\Delta'} D(s)$ versus $s/\Delta z^{\lambda'}$. We vary $\Delta'$ to find the value for which $D(s)$ approaches a plateau and vary $\lambda'$ to find the value that collapses the low-$s$ crossovers. We find a flat plateau and collapsing crossovers for $\Delta' = 0.47(5)$ and $\lambda' = 1.9(1)$.  Results are shown in Fig.~\ref{fig:DOS_rel}b. Note that $\Delta'$ and $\lambda'$ are in good agreement with the critical exponents $\Delta$ and $\lambda$, respectively, that govern response. Below we will show that $\Delta' = \Delta$ and $\lambda' = \lambda$.

To better understand the nature of the abundant low-$s$ modes in the relaxational density of states, we relate their eigenrate to the non-affine character of their motion. 
The prevalence of sliding over normal relative motion can be quantified by the ratio
\begin{equation}
\Gamma_n^2 \equiv \left. \frac{ \overline{(\Delta u^\perp)^2} } {\overline{(\Delta u^\parallel)^2} } \right |_n \,,
\label{eqn:Gamma}
\end{equation}
which is evaluated for the relative displacements of mode $n$. Bars indicate averages over all contacts. We now show that $\Gamma_n$ becomes larger as $s_n$ is decreased, indicating the growing importance of sliding motion.

\begin{figure}[tbp]
\centering
\includegraphics[clip,width=0.9\linewidth]{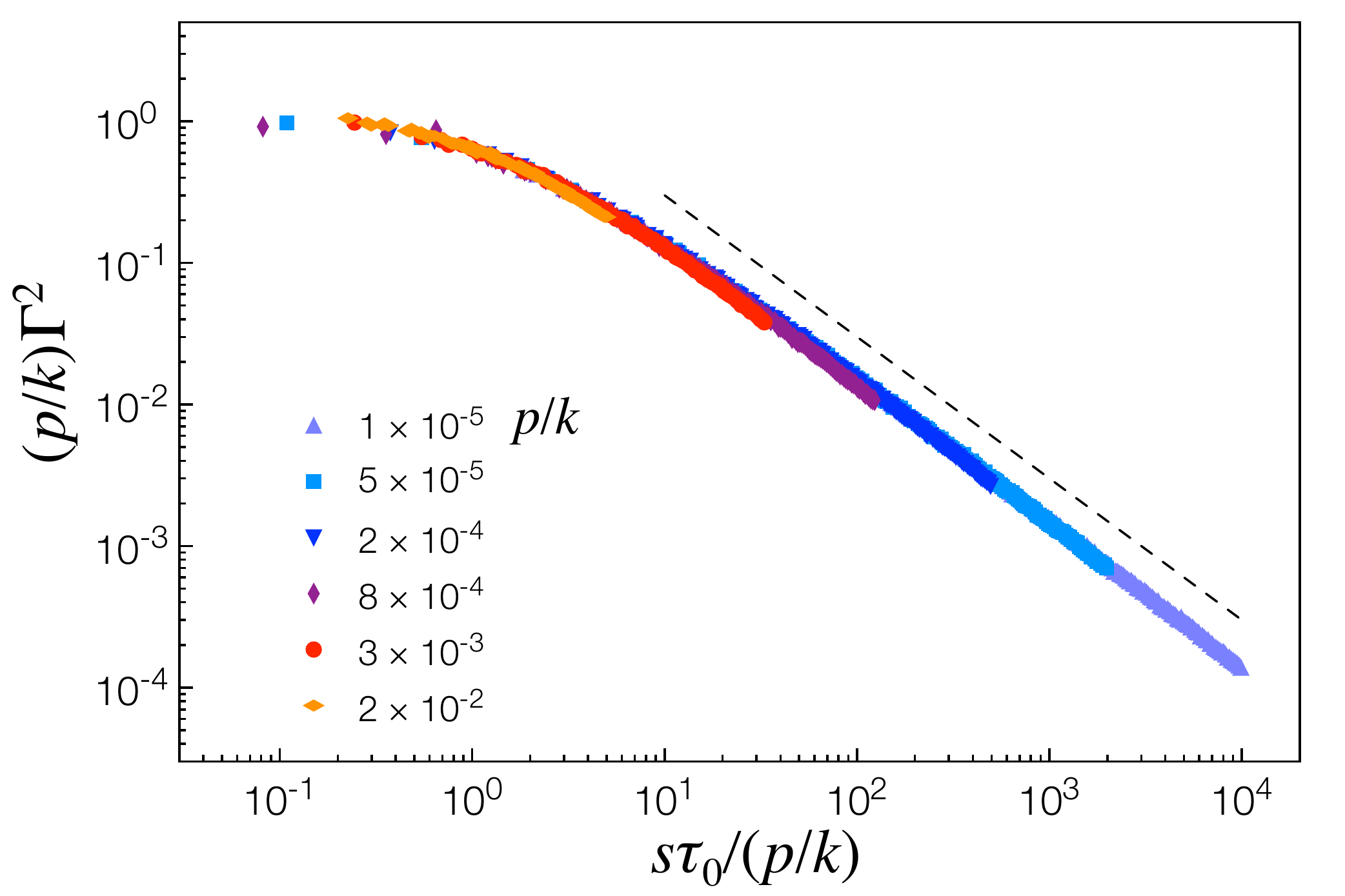}
\caption{The ratio $\Gamma = \Delta u^\perp/\Delta u^\parallel$ as a function of relaxational eigenrate $s$ collapses when rescaled with confining pressure $p$ according to Eq.~(\ref{eqn:nonaff}). The dashed line has slope one.}
\label{fig:nonaff}
\end{figure}

Each eigenrate of the material satisfies 
\begin{equation}
s_n = \frac{ \langle s_n | \hat {\cal K} | s_n \rangle }{ \langle s_n | \hat {\cal B} | s_n \rangle} 
   = \left. \frac{U - U^0}{R}\right |_n \,,
\label{eqn:UR}
\end{equation}
where $U$ and $R$ are the elastic potential energy and Rayleigh dissipation function of Eqs.~(\ref{eqn:V}) and (\ref{eqn:R}), respectively. By considering typical values of the relative displacements, Eq.~(\ref{eqn:UR}) relates $s_n$ to $\Gamma_n$. Dropping subscripts, the relation is
\begin{equation}
s \tau_0 \sim \frac{\Gamma^{-2} + c (p/k)}{\Gamma^{-2} + c' \beta } \,,
\label{eqn:dispersion00}
\end{equation}
where the prefactors $c$ and $c'$ are constants expected to be of order unity. To reach Eq.~(\ref{eqn:dispersion00}), we have used the fact that the typical force in the reference state is proportional to the pressure and  have  assumed that $\Delta u^\perp$ dominates sliding at the contact.
Specializing to the case $\beta = 1$, where normal and tangential motion are damped equally, and solving for $\Gamma$ in the limit of small $s$ gives
\begin{equation}
\frac{1}{ \Gamma^2} \sim \left( \frac{p}{k} \right)\left[{ {\rm const} + \frac{s\tau_0}{(p/k)} } \right]\,.
\label{eqn:nonaff}
\end{equation}
Thus the slower the mode, the stronger the dominance of sliding over normal motion. Eq.~(\ref{eqn:nonaff}) is verified in Fig.~\ref{fig:nonaff}, which plots $\Gamma$ for six packings (no ensemble averaging) at different confining pressures. The pre-stress introduces an upper bound on the non-affinity; $\Gamma$ approaches a constant for relaxation rates $s \lesssim (p/k)/\tau_0$, which coincides with $s^*$.  At the crossover rate $s^*$ sliding exceeds normal motion by a factor $\Gamma^* \sim 1/\Delta z$.

\subsection{Predicting the relaxational density of states}
The exponents $\Delta'$ and $\lambda'$ characterize the form of the density of states. We now seek to determine them analytically. Several years ago, Wyart, Nagel, and Witten (WNW) used a variational method \cite{wyart05} to explain the anomalous plateau in the vibrational density of states $D(\tilde{\omega})$ -- the probability density of vibrational eigenfrequencies $\tilde{\omega}$ of packings { with} inertia and { without} damping. For the case where all particles have unit mass, the square of each eigenfrequency, $\tilde \omega_n^2$, is an eigenvalue of the stiffness matrix $\hat {\cal K}$. The undamped vibrational modes are therefore a natural basis in which to express {\em quasistatic} deformations.

The work of WNW has become a landmark of the jamming literature. Here we generalize the method to { overdamped} dynamics and show that it correctly predicts the values of $\Delta'$ and $\lambda'$. Without going into technical details, we sketch the key ideas of the method and show how they can be extended to overdamped dynamics.

The central idea of the WNW method is as follows. One takes a packing of particles with contact number $z = z_c + \Delta z$ and lays a mesh over it. The packing has linear size $L$ and the mesh has a characteristic length $q^{-1}$. Everywhere a contact crosses a face of the mesh, that contact is ``cut'' or ignored. If enough contacts are cut, the packing loses rigidity and zero energy collective motions, i.e.~floppy modes, appear. For this to happen the number of cut contacts, which is on the order of the total surface are of the mesh ${ O}(q\, \Omega)$, must exceed the number of contacts the original packing had in excess of the isostatic value, which is ${ O}(\Delta z \, \Omega)$. Therefore floppy modes appear whenever
$q > q^* \sim \Delta z $. The diverging length scale $1/q^*$ is known as the isostatic length \cite{tkachenko99,wyart05}. When $q > q^*$ the cutting process creates on the order of $(q-q^*) \Omega$ floppy modes. 

The floppy modes can be exploited to construct trial modes $\lbrace |Q_{\rm WNW} \rangle \rbrace$ for use in a variational argument. Each trial mode is a floppy mode modulated by a sinusoidal envelope that vanishes on the faces of the mesh. This ``repairs'' the large relative motions between particles participating in cut contacts. 

Having determined how to construct trial modes, the vibrational density of states can be determined in two steps. Noting that trial modes are parameterized by their ``wavenumber'' $q$, one first determines the probability density $D(q)$. The integral of $D(q)$ is related to the number density of trial modes, $\int_{q^*}^q {\rm d}q' \, D(q') \sim q-q^*$.
Differentiating with respect to $q$ gives
\begin{equation}
D(q) \sim {\rm const} \,\,\,\,\,\, {\rm for}  \,\,\,\,\,\, q \gtrsim q^* \,.
\end{equation} 
The second step is to relate the trial mode wavenumber $q$ to the frequency $\tilde{\omega}$ via a dispersion relation $\tilde{\omega} = \tilde{\omega}(q)$; one then has
$D(\tilde{\omega}) = D(q)|{\rm d}q/{\rm d}\tilde{\omega}| \sim |{\rm d}q/{\rm d}\tilde{\omega}|$. For undamped modes the dispersion relation turns out to be linear in $q$, so $D(\tilde \omega)$ develops a plateau.

For our purposes, the key observation is that the WNW argument is exclusively geometric in nature, up to the point where the dispersion relation $\tilde{\omega}(q)$ is invoked. The undamped nature of the dynamics enters only in this final step. Recall that the undamped vibrational modes are the natural basis with which to describe quasistatic deformations. We anticipate that trial modes that work well for quasistatic deformations will also work well for sufficiently slow deformations. Therefore we can use the same WNW trial modes, parameterized by $q$, and apply them to overdamped dynamics to estimate the relaxational density of states,
\begin{equation}
D(s) = D(q) |{\rm d}q/{\rm d}s| \sim |{\rm d}q/{\rm d}s|  \,\,\,\,\,\, {\rm for}  \,\,\,\,\,\, s \gtrsim s^* \,.
\label{eqn:Ds}
\end{equation}
Here we have introduced the crossover rate $s^* \equiv s(q^*)$, which we identify with $1/\tau^*$.

It remains only to determine the overdamped dispersion relation $s = s(q)$. To do this we exploit Eq.~(\ref{eqn:dispersion00}), the relation between relaxation rates and non-affinity. Recall that trial modes are constructed by applying a sinusoidal modulation to a floppy mode created by the cutting procedure. Because of the modulation, the relative normal and tangential motions of the trial mode differ from those in the floppy mode. As the sinusoid has wavenumber $q$, its contribution to the relative motions is of order $q \, u_{\rm fm}$, where $u_{\rm fm}$ is  the typical displacement in the floppy mode. This amounts to a small correction to the tangential motions, which are nonzero in the floppy mode. However, it sets the scale of the relative normal motions of the trial mode, because the relative normal motions of the floppy mode are zero, to leading order. Therefore $\Gamma_{\rm WNW} = \Delta u_{\rm WNW}^\perp / \Delta u_{\rm WNW}^\parallel \sim 1/q$.

The dispersion relation for trial modes is thus
\begin{equation}
s \tau_0 \sim \frac{q^2 + c (p/k)}{ q^2 + c' \beta } \,.
\label{eqn:dispersion0}
\end{equation}
Recalling that $q > q^*$ for all trial modes, the dispersion relation can be expanded, keeping only the dominant term:
\begin{equation}
s \tau_0 \sim \left \lbrace 
	\begin{array}{cc}
	q^2/\beta &\,\,\,\,\,\,\,\,\,\,\,\,\, q \ll \beta^{1/2} \\
	{\rm const} & \beta^{1/2} \ll q \,.
	\end{array} \right.
\label{eqn:disp}
\end{equation}
As the the dispersion relation is independent of $p$ to leading order, the pre-stress can be neglected.
When normal and tangential motions are damped similarly, $\beta$ is of order unity and the dispersion relation is quadratic. That overdamped dynamics should display a quadratic dispersion relation is not surprising; consider, e.g., plane waves in an overdamped ball-and-spring chain. We stress, however, that this quadratic form was not assumed, but derived from properties of the trial modes. Indeed, when the damping ratio $\beta \lesssim (q^*)^2$, i.e.~when sliding motion is weakly damped,  $q$ is always larger than $\beta^{1/2}$, the dispersion relation is no longer quadratic, and there is a crossover to qualitatively different response. We return to this observation in Section IV.

For later convenience we write the $q \ll \beta^{1/2}$ dispersion relation as
\begin{equation}
s \tau_0 \sim  \, \frac{q^{\lambda'}}{\beta}  \,\,\,\,\,\,\,\, {\rm for }  \,\,\,\,\,\,\,\,  s \gtrsim s^* \,.
\label{eqn:dispersion}
\end{equation}
Hence $\lambda' = 2$ and 
\begin{equation}
s^* \tau_0 = \frac{(q^*)^{\lambda'}}{\beta } \sim \frac{\Delta z^{\lambda'}}{\beta}  \,.
\end{equation}
Invoking Eq.~(\ref{eqn:Ds}), the density of states scales as
\begin{equation}
\frac{D(s)}{\beta\tau_0} \sim  \left( \frac{1}{\beta s\tau_0} \right)^{\Delta'} \,\,\,\,\,\, {\rm for}  \,\,\,\,\,\,  s \gtrsim s^*  \,,
\label{eqn:dos}
\end{equation}
with $\Delta' = 1/{\lambda'} = 1/2 $. The values of both $\lambda'$ and $\Delta'$ are in excellent agreement with the numerical results of Section \ref{sec:DOS} for $\beta = 1$. We will confirm the scaling with $\beta$ in Section \ref{sec:weak}.

\subsection{Predicting the complex shear modulus}
With the form of the relaxational density of states in hand, we can now return to the complex shear modulus. We restrict our attention to the case $\beta = 1$ until indicated otherwise. We shall show that the divergence in $D(s)$ and the crossover rate $s^*$, through the exponents $\lambda'$ and $\Delta' $, suffice to predict the scaling of $G^*(\omega)$ demonstrated above.  Specifically, we will show that $\mu = \lambda' \Delta' $, $\lambda = \lambda' $, and $\Delta = \Delta' $.

Eq.~(\ref{eqn:compliance}) relates the complex compliance to an integral over modes. Using our results for the relaxational density of states, the integrals of Eq.~(\ref{eqn:compliance}) can be made into bounds on the scaling of $J^*$ by replacing $\int_0^\infty ( \cdot ) D(s) \,{\rm d }s$ with $ \tau_0 \int_{s^*}^\infty (\cdot) (s\tau_0)^{-\Delta'}  {\rm d}s$.
We shall assume that the bounds are saturated, which is verified by the good agreement between the resulting predictions and the numerical results presented above.

{\em Quasistatic response.---} We begin with the zero frequency or quasistatic limit. The quasistatic shear modulus $G_0$ and dynamic viscosity $\eta_0$ of a viscoelastic solid obey
$1/G_0 = \lim_{\omega \rightarrow 0} J'(\omega) $ 
and 
$\eta_0/G_0^2 = - \lim_{\omega \rightarrow 0 }[ {J''(\omega)}/{\omega} ]$. 
Eqs.~(\ref{eqn:compliance}), (\ref{eqn:dispersion}), and (\ref{eqn:dos}) give $G_0$:
\begin{equation}
\frac{k}{G_0} \sim \frac{1}{\tau_0^{\Delta'}} \int_{s^*}^{1/\tau_0}  \frac{{\rm d}s}{s^{1 + \Delta'}} \sim  \frac{1} {(q^*)^{\lambda' \Delta'}}  \,.
\label{eqn:G0int}
\end{equation}
Using $q^* \sim \Delta z $,  it immediately follows that the quasistatic shear modulus scales as 
$G_0 \sim k \, \Delta z^{\mu} $
with $\mu = {\lambda' \Delta'} = 1$. This explains the numerical finding of Eq.~(\ref{eqn:G0}) and is in good agreement with prior numerics \cite{ohern03,ellenbroek06}.
Our result is compatible with previous calculations of the quasistatic shear modulus \cite{zaccone11,wyart05c} and is, to our knowledge, the first to directly relate $G_0$ to the isostatic length $1/q^*$.

Proceeding in the same way, one finds that the dynamic viscosity scales as
$ \eta_0 \sim G_0 \tau^* \sim {k\tau_0}/{\Delta z^{ \lambda' - \mu }} $ with exponent $\lambda' - \mu  = 1$.
Hence $\lambda = \lambda'$,  as anticipated, and we have also explained the numerical finding of Eq.~(\ref{eqn:eta0}).

{\em Shear thinning regime.---} For frequencies $s^* \lesssim \omega \lesssim 1/\tau_0$, quasistatic predictions fail and the response becomes shear thinning. The real and imaginary parts of the complex compliance are
\begin{eqnarray}
k\,{J'} &\sim& 
\frac{1}{\tau_0^{\Delta'}} 
\left[\int_{s^*}^\omega \frac{s^{1- \Delta'}\,{\rm d}s}{\omega^2}
+ 
\int_\omega^{1/\tau_0}  \frac{{\rm d}s}{s^{1 + \Delta'}} \right]  \nonumber \\
&\sim &
\left(\frac{1}{\omega \tau_0}\right)^{\Delta'} \,,
\label{eqn:thinningJp}
\end{eqnarray}
and
\begin{eqnarray}
k\,{J''} &\sim& 
\frac{1}{\tau_0^{\Delta'}} 
\left[\int_{s^*}^\omega \frac{{\rm d}s}{\omega \, s^{ \Delta'}}
+ 
\int_\omega^{1/\tau_0}  \frac{{\omega \, \rm d}s}{s^{2 + \Delta'}} \right]  \nonumber \\
&\sim &
\left(\frac{1}{\omega \tau_0}\right)^{\Delta'} \,.
\label{eqn:thinningJpp}
\end{eqnarray}
It immediately follows that 
$G^* \sim k\, (\imath \omega\tau_0)^{\Delta} $ with $\Delta = \Delta'  = 1/2$,
again in excellent agreement with numerics.

In linear response the stress relaxation modulus $G_r(t)$, which gives the stress response after a unit step strain at time $t = 0$, can be calculated directly from the complex shear modulus. The scaling $G^* \sim (\imath \omega)^\Delta$ implies a regime of power law relaxation $G_r(t) \sim t^{-\Delta}$. Therefore our results explain the critical relaxation $G_r \sim t^{-0.5}$ seen in simulations of athermal suspensions near jamming \cite{hatano09}.

{\em High frequency response.---} The density of states vanishes above an upper bound on the order of $1/\tau_0$. For driving frequencies that exceed the bare rate, one finds
$J'  \sim {1}/{k(\omega \tau_0)^2}$ and $J'' \sim -{1}/{k \omega \tau_0}$. 
Hence the high frequency moduli are simply given by the coefficients of the microscopic force law,
\begin{equation}
G' \sim k \,\,\,\,\,\,\, {\rm and} \,\,\,\,\,\,\, G'' \sim k \omega \tau_0 \,.
\end{equation}
This is consistent with the observation that high frequency response is affine.

To summarize, when expressed in terms of $\Delta \phi$, the storage modulus scales as
\begin{equation}
G'  \sim 
\left \lbrace \begin{array}{cc}
k\, \Delta \phi^{1/2} & \,\,\,\,\,\,\,\,\,\,\,\,\,\,\,\,\,\,\,\,\,\,\, \omega \tau_0 < \Delta \phi  \\
k \, (\omega \tau_0)^{1/2} & \, \Delta \phi  < \omega \tau_0 < 1 \\
k & 1  < \omega \tau_0 \,, 
\end{array} \right.
\end{equation}
and the loss modulus scales as
\begin{equation}
G''  \sim 
\left \lbrace \begin{array}{cc}
k(\omega \tau_0)  / \Delta \phi^{1/2} & \,\,\,\,\,\,\,\,\,\,\,\,\,\,\,\,\,\,\,\,\,\,\, \omega \tau_0 < \Delta \phi  \\
k \, (\omega \tau_0)^{1/2} &  \, \Delta \phi  < \omega \tau_0 < 1 \\
k (\omega \tau_0) & 1  < \omega \tau_0  \,.
\end{array} \right.
\end{equation}

\begin{figure}[tbp]
\centering
\includegraphics[clip,width=0.9\linewidth]{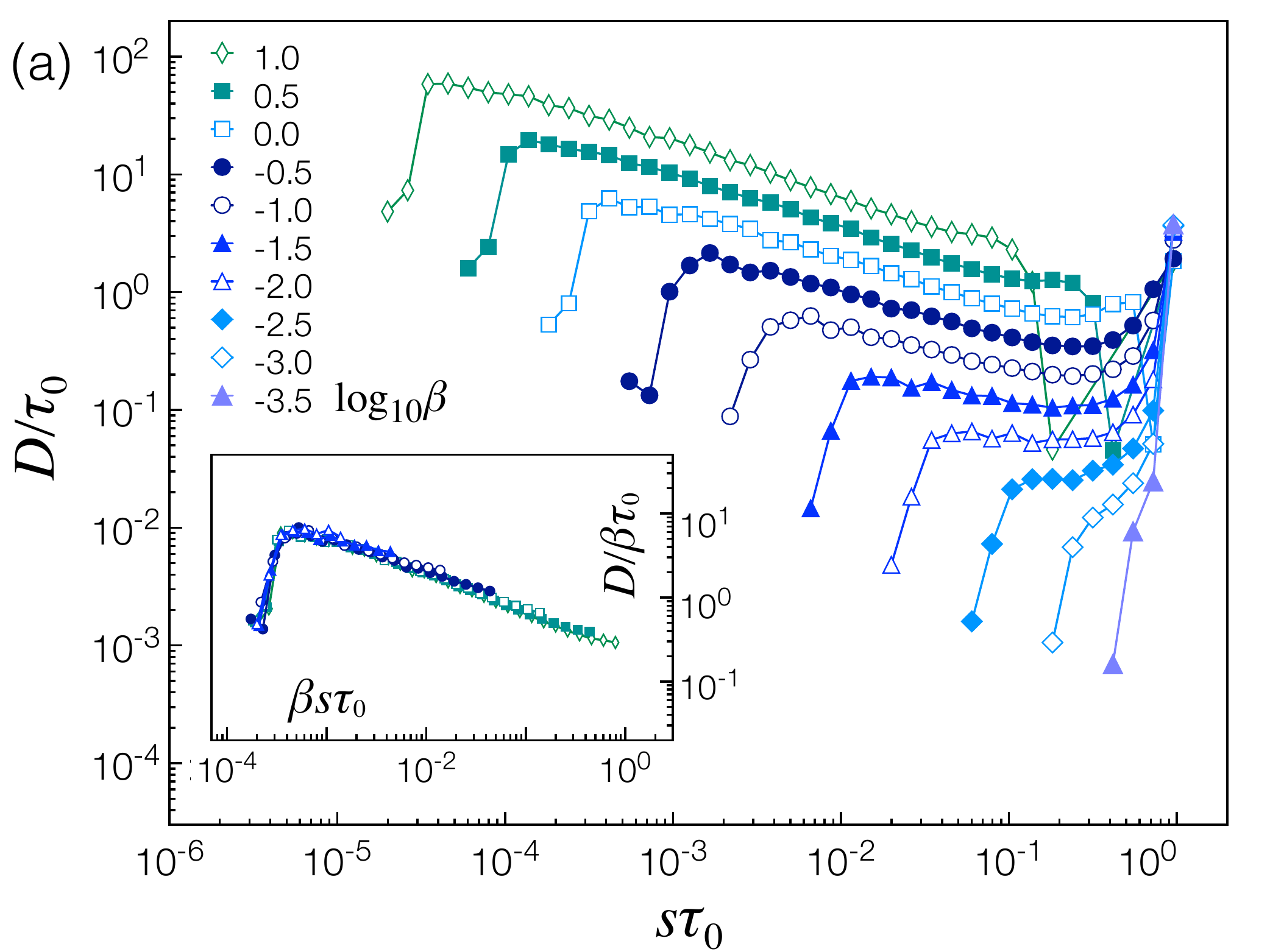}\\
\vspace{0.4cm}
\includegraphics[clip,width=0.9\linewidth]{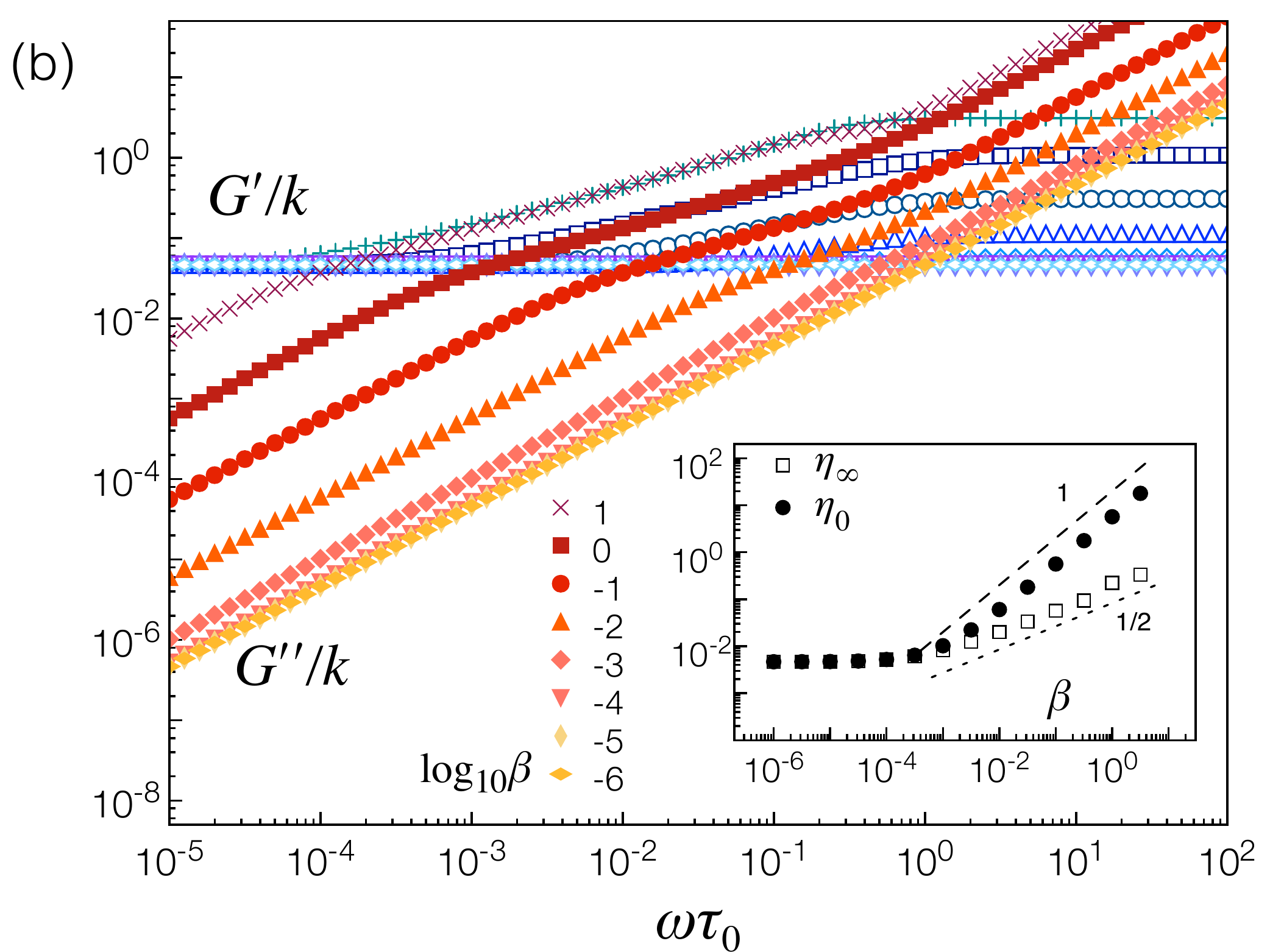}
\caption{
(a) Relaxational density of states for $\Delta z = 0.063$ and varying damping ratio $\beta$ (legend). (b) Evolution of the relaxational density of states for $\Delta z = 0.063$ and varying $\beta$.  Inset: Dynamic ($\omega \rightarrow 0$) viscosity $\eta_0$  and affine ($\omega \rightarrow \infty$) viscosity $\eta_\infty$ as a function of $\beta$.
}
\label{fig:beta0}
\end{figure}

\section{Weakly damped sliding}
\label{sec:weak}

In the preceding Sections the damping ratio $\beta $ was taken to be unity, so relative normal and tangential motions were damped with equal strength. The dispersion relation, Eq.~(\ref{eqn:disp}), then has a simple quadratic form. This is significant because the form of the relaxational density of states and the complex shear modulus both follow from the dispersion relation. As the  dispersion relation ceases to be quadratic when sliding  is weakly damped, one anticipates changes in $D(s)$ and $G^*(\omega)$. We therefore investigate the limit of weak tangential damping.

The $\beta \rightarrow 0$ limit is important for an additional reason. While it may seem obvious that sliding motions are damped in materials such as foams and emulsions,  many numerical studies take $\beta $ to be zero, so that only relative normal motions dissipate energy \cite{roux08,hatano08,otsuki09,heussinger10}. This is done for reasons of numerical convenience: with sliding undamped, neither elastic nor viscous forces impose torques, so the particles' rotational degrees of freedom can be ignored. We will show that this choice of force law qualitatively alters a system's viscoelastic response.

The condition for weakly damped sliding is $\beta \ll \Delta z^2$.
When sliding motion is weakly damped, the dispersion relation $s \tau_0 \sim {\rm const}$ is independent of $q$ to leading order -- see Eq.~(\ref{eqn:disp}). This already suggests a dramatic departure from the dynamic critical scaling described above, as sending $q \rightarrow q^*$ no longer produces a diverging time scale.

To see how response evolves as a system passes from strong to weak tangential damping, in Fig.~\ref{fig:beta0}a
we plot the relaxational density of states  for varying $\beta$ averaged over 20 packings with excess coordination $\Delta z = 0.063$. Several features are apparent. For $\beta > 1$ there is a gap between slow modes with $\beta s \tau_0 \lesssim  O(1)$, which has the square root divergence discussed above, and a narrow band with $s\tau_0 \sim O(1)$. 
$D(s)$ narrows with decreasing $\beta < 1$, consistent with the form of the dispersion relation. This is because the crossover rate $s^* \sim \Delta z^2/\beta \tau_0$ increases, and in so doing ``squeezes out'' the anomalous slow modes: the interval over which $D(s)$ displays $1/s^\Delta$ growth becomes increasingly narrow. In the inset of Fig.~\ref{fig:beta0}a we plot $D/\beta \tau_0$ versus $\beta s \tau_0 $ for slow modes in the strong tangential damping regime ($s\tau_0  < 0.1$ and $\beta > \Delta z^2$). The data collapse is excellent, confirming the $\beta$ dependence of Eq.~(\ref{eqn:dos}).

In Fig.~(\ref{fig:beta0}b) we plot the complex shear modulus for varying $\beta$. Clearly the form of $G^*$ is dramatically different when sliding is weakly damped. In particular, the critical shear thinning regime narrows as $\beta \rightarrow 0$ and vanishes when $s^*\tau_0 \sim O(1)$, which coincides with the weak tangential damping regime. Both the dynamic viscosity $\eta_0 \sim \beta$ and the high frequency (affine) viscosity $\eta_\infty \sim \beta^{1/2}$ depend on the damping ratio when sliding is strongly damped (Fig.~\ref{fig:beta0}b inset), but become independent of $\beta$ in the weak tangential damping regime.
To understand the crossover in $\eta_0$, recall that when sliding is strongly damped, the dynamic viscosity $\eta_0 \sim (\Delta u^\perp/\gamma)^2 \sim \beta/\Delta z$ reflects the scaling of the typical relative tangential displacement in quasistatic response. When sliding is weakly damped, the dominant source of dissipation is relative normal motion and $\eta_0 \sim (\Delta u^\parallel/\gamma)^2$, independent of $\beta$. Ellenbroek et al.~find $\Delta u^\parallel/\gamma \sim \Delta z^{1/2} $ in the quasistatic limit \cite{ellenbroek06}. 

To gain further insight into the weak tangential damping regime, it is useful to directly consider the case $\beta = 0$ while also neglecting the pre-stress. The pre-stress can be neglected because it does not appear in the leading order term in the dispersion relation. The key observation is that under these conditions, the damping matrix $\hat {\cal B}$ and the stiffness matrix $\hat {\cal K}$ are proportional, $\hat {\cal B} = \tau_0 \hat {\cal K}$. The equation of motion becomes
\begin{equation}
( 1 + \imath \omega \tau_0  ) \hat {\cal K} |Q(\omega) \rangle = \sigma(\omega) |\hat \sigma \rangle \,.
\label{eqn:proportional}
\end{equation}

By expanding $|Q(\omega)\rangle$ in the undamped eigenmodes of the stiffness matrix $\hat {\cal K}$, Eq.~(\ref{eqn:proportional}) can be solved for the complex shear modulus. The result is
\begin{equation}
G^*(\omega) = G_0 \left( 1 + \imath  \omega \tau_0 \right) \,.
\label{eqn:beta0}
\end{equation}
The quasistatic  modulus $G_0 \sim k \, \Delta z$ is unchanged, but the dynamic viscosity $\eta_0 = G_0 \tau_0$ now { vanishes}, as anticipated from the quasistatic scaling of $\Delta u^\parallel$. The time scale $ \eta_0 / G_0 $, which equals $\tau^*$ when sliding is strongly damped, no longer diverges. Moreover, consistent with our numerical observations, there is no critical shear thinning regime.

The above discussion establishes an important intuition. Quasistatic response near unjamming is strongly non-affine \cite{ellenbroek06,wyart08}, reflected in the critical divergence of sliding motion on approach to the unjamming point. 
When non-affine motion dissipates energy, its divergence is reflected in the viscosity; see also Refs.~\cite{tighe11,heussinger12,tighe12,during12}. Dynamic critical scaling, which describes response at finite rates, requires a diverging time scale. We have shown that non-affine motion must be strongly damped to have a diverging time scale in viscoelastic linear response. When sliding is damped weakly or not at all, there is no diverging time scale and hence no dynamic critical response, reflected in the absence of a shear thinning regime.

\section{Conclusions}

Linear response in viscoelastic jammed solids displays dynamic critical scaling when sliding is strongly damped. We have shown that critical response is driven by slow relaxational eigenmodes. Close to unjamming the slow modes are anomalously abundant and have a spatially non-affine character reminiscent of floppy modes. By extending the WNW method to overdamped dynamics, the broad distribution of relaxation rates can be explained. The complex shear modulus can then be derived from the relaxational density of states.

Our results clearly establish that nonlinear viscous forces are not necessary for a material to be  shear thinning and, concomitantly, that low frequency rheology near jamming does not trivially reflect the form of the microscopic viscous force law. We have also shown that viscoelastic response displays a sensitive dependence on the form of the viscous force law. The critical shear thinning regime only exists when sliding motion is strongly damped, and  in the weak tangential damping regime the dynamic viscosity vanishes on approach to unjamming. We consider it likely that the qualitative distinction between strong and weak tangential damping persists in nonlinear response, including steady flow. Therefore simulations that damp only normal motion \cite{roux08,hatano08,otsuki09,heussinger10} may not be comparable with those that include damping of tangential motion \cite{tighe10c,sexton11,seth11}.

Our calculations can be compared with a number of experimental results \cite{liu96,cohen-addad98,gopal03,krishan10,kropka10,lietor-santos11}. As noted above, the quasistatic regime has yet to be observed. Foams and emulsions, for example, are not thermodynamically stable; their structure evolves due to coarsening and related effects not included in the bubble model. As a result, their low-frequency rheology depends on the age of the sample \cite{hohler05}. More encouragingly, there is experimental evidence of $\omega^{0.5}$ scaling in the complex shear modulus of emulsions \cite{liu96}, liquid foams \cite{cohen-addad98,gopal03,krishan10}, organic foams \cite{kropka10}, and soft suspensions \cite{lietor-santos11}, consistent with our prediction for $G^*$ in the critical regime. 
This comparison must be made with caution, however, because the viscous force law is not always known and likely varies from material to material. While viscous forces are likely to be linear in some of these materials, the bubble-bubble drag force in foams is known to be nonlinear \cite{katgert08,bretherton,denkov09}. An important goal for future research will be to identify the full dependence of macroscopic rheology near jamming on the form of the microscopic force laws, including nonlinear viscous forces.

\begin{acknowledgements}
This work was supported by the Dutch Organization for Scientific Research (NWO).
\end{acknowledgements}

\bibliographystyle{apsrev}
\bibliography{tighe}

\end{document}